\documentclass[twocolumn,amsmath,amssymb]{revtex4}
\usepackage{graphicx}
\usepackage{dcolumn}
\usepackage{bm}

\begin{document}


\title{An automated and versatile ultra-low temperature SQUID magnetometer}
\author{A. Morello}
\affiliation{Kamerlingh Onnes Laboratory, Leiden Institute of
Physics, Leiden University, P.O. Box 9504, NL-2300 RA
 Leiden}
\author{W. G. J. Angenent}
\affiliation{Kamerlingh Onnes Laboratory, Leiden Institute of
Physics, Leiden University, P.O. Box 9504, NL-2300 RA
 Leiden}
\author{G. Frossati}
\affiliation{Kamerlingh Onnes Laboratory, Leiden Institute of
Physics, Leiden University, P.O. Box 9504, NL-2300 RA
 Leiden}
\author{L. J. de Jongh$^*$}
\affiliation{Kamerlingh Onnes Laboratory, Leiden Institute of
Physics, Leiden University, P.O. Box 9504, NL-2300 RA
 Leiden}


\date{\today}

\begin{abstract}
We present the design and construction of a SQUID-based
magnetometer for operation down to temperatures $T \simeq 10$ mK,
while retaining the compatibility with the sample holders
typically used in commercial SQUID magnetometers. The system is
based on a \textit{dc}-SQUID coupled to a second-order
gradiometer. The sample is placed inside the plastic mixing
chamber of a dilution refrigerator and is thermalized directly by
the $^3$He flow. The movement though the pickup coils is obtained
by lifting the whole dilution refrigerator insert. A
home-developed software provides full automation and an easy user
interface.

\end{abstract}

\maketitle

\section{Introduction}
The use of Superconducting QUantum Interference Devices (SQUIDs)
\cite{gallopB} in ultra-sensitive magnetic measurements systems
may nowadays be considered as a standard technique, to the extent
that several companies offer reliable and automated commercial
SQUID magnetometers. However, no commercial magnetometer has yet
become available for operation at millikelvin temperatures
utilizing a dilution refrigerator. A few systems have been
constructed that combine the sensitivity of SQUID magnetometery
with the millikelvin temperature range attainable by means of
$^3$He/$^4$He dilution refrigerators
\cite{paulsenB,wernsdorfer00JAP(SQUID),bjornsson01RSI}, at the
cost of requiring the sample to be mounted on dedicated sample
holders, which ensure the thermalization to the cold plate of the
refrigerator by means of conducting thermal links.

We describe here a SQUID magnetometer designed to allow
high-sensitivity measurements down to $T \sim 10$ mK, while
retaining the compatibility with the sample holders used in
commercial magnetometers. Moreover, we wanted to avoid building a
system dedicated exclusively to SQUID magnetometery. Our setup can
accommodate several different experimental probes, e.g. for
nuclear magnetic resonance \cite{morello04CM},
inductance-bridge-based \textit{ac}-susceptometery
\cite{morello03PRL} and resistivity measurements. Finally, we
wrote a software that fully automates the system and offers a very
user-friendly interface to program the measurement sequences and
analyze the data.

An example of research for which our setup was particularly
designed is the field of metallic nanoclusters, which show
spectacular quantum-size effects in their thermodynamic properties
\cite{volokitin96N} at very low temperatures, as a consequence of
which their magnetic susceptibility is very small in this regime.
Furthermore, since most of these materials are highly
air-sensitive, it is a necessity to introduce and keep the sample
in a sealed glass tube, that allows at the same time to perform
preliminary measurements at $T>1.8$ K in commercial SQUID
magnetometers and pre-check the quality of the sample in a fast
and cheap way. The sample, in its sample holder, can thereupon be
inserted in the ultra-low temperature SQUID magnetometer without
need of any further manipulation, which constitutes an obvious
advantage.

The really challenging part of the design consists in the fact
that, like in most commercial and non-commercial
\cite{nave80RSI,genna91RSI} magnetometers, to obtain the absolute
value of the magnetization we need to move the sample through a
gradiometer pick-up coil, but in our case the sample is at $T \geq
10$ mK inside the mixing chamber of the dilution refrigerator! The
sample movement is needed because the coils of the gradiometer
will never perfectly compensate one another, leaving an empty-coil
signal that will add a spurious contribution to the measured
magnetic moment, or even wash out the signal of the sample to be
measured. As we shall discuss below, this has required a system
that moves the \emph{whole} dilution refrigerator insert.

\section{Dilution refrigerator} \label{sec:dilution}

\begin{figure}[t]
\includegraphics[width=8.5cm]{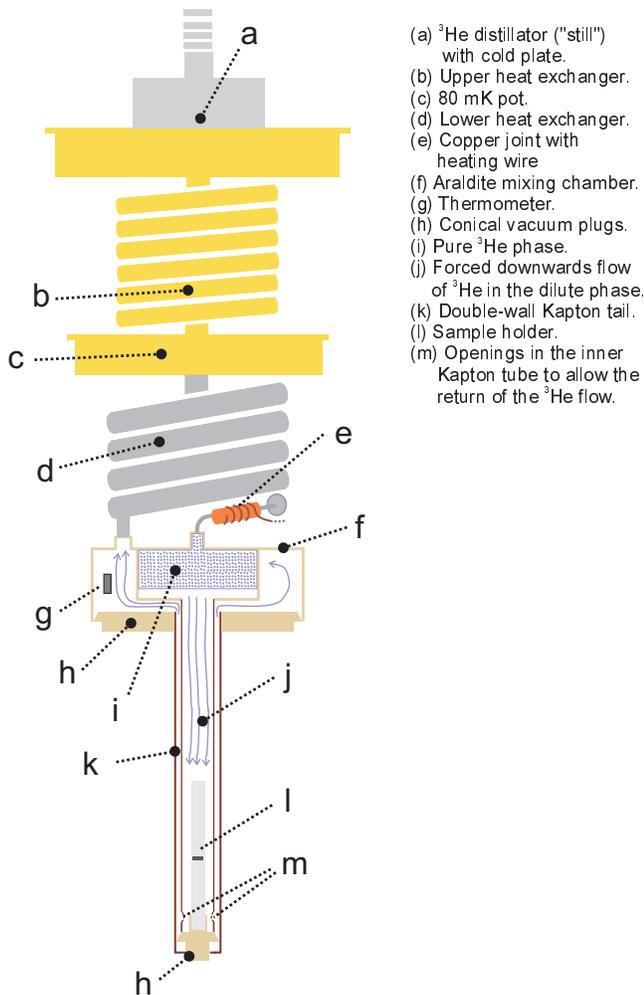}
\caption{\label{dilution} Scheme of the low-$T$ part of the
dilution refrigerator. For clarity we omit the vacuum can and the
radiation shields anchored at the still and at the 80 mK pot. Only
the narrow tail of the refrigerator is inserted in the bore of a 9
T superconducting magnet (not shown).}
\end{figure}
Our ultra-low temperature setup is based on a Leiden Cryogenics
MNK126-400ROF dilution refrigerator. Instead of the standard
copper mixing chamber, we fitted the system with a specially
designed plastic mixing chamber that allows the sample to be
thermalized directly by the $^3$He flow. A scheme of the
low-temperature part of the refrigerator is shown in Fig. 1. The
mixing chamber consists of two concentric tubes, obtained by
rolling a Kapton foil coated with Stycast 1266 epoxy. The tops of
each tube are glued into concentric Araldite pots: the inner pot
receives the downward flow of condensed $^3$He and, a few
millimeters below the inlet, the phase separation between pure
$^3$He phase and dilute $^3$He/$^4$He phase takes place. The
circulation of $^3$He is then forced downwards along the inner
Kapton tube, which has openings on the bottom side to allow the
return of the $^3$He stream through the thin space in between the
tubes. Both the bottom of the Kapton tail and the outer pot are
closed by conical Araldite plugs smeared with Apiezon N grease.
The sample is typically placed inside a capsule filled with cotton
and inserted in a plastic straw, fixed onto the bottom Aradite
plug. Alternatively, for samples being air sensitive or having
very small magnetic signal, one may largely reduce the background
contribution of the holder by placing the sample between two
Suprasil glass rods inside a Suprasil tube \cite{sinzigT}, as
shown in Fig. 2. In this way, the sample is placed in a symmetric
configuration with respect to the pick-up coil systems, so that
the only signal induced by the vertical movement is that
originating from the sample space and the sample itself. Both
types of sample holders are compatible with the commercial SQUID
magnetometers (Cryogenic Consultant Limited S600c, Quantum Design
MPMS-5 and MPMS-XL) available in our laboratory for magnetic
measurements at $T>1.8$ K.

\begin{figure}[t]
\includegraphics[width=8.5cm]{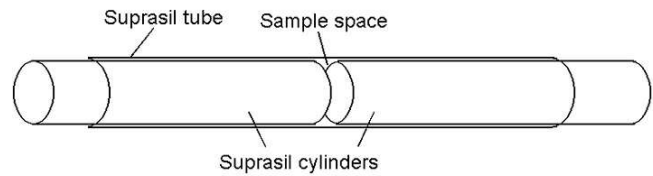}
\caption{\label{suprasil} Sketch of the symmetric Suprasil glass
sample holder.}
\end{figure}

The Kapton tail, which is about 35 cm long, is surrounded by two
silver-plated brass radiation shields, one anchored at the 80 mK
pot, the other at the still. The whole low-$T$ part of the
refrigerator is closed by a vacuum can, which has itself a thin
brass tail to surround the Kapton part of the mixing chamber, and
is inserted into the bore of an Oxford Instruments 9 Tesla NbTi
superconducting magnet. In this way, only Kapton and brass
cylinders (plus the lowest Araldite plug) are placed in the
high-field region, whereas all other highly-conducting metal parts
(heat exchangers, 80 mK pot, $^3$He distillator, etc.) are outside
the magnet bore and subject only to a small stray field. This
design minimizes the eddy current heating resulting from moving
the refrigerator through the pick-up coil of the SQUID
magnetometer, while at the same time thermalizing the sample
directly by the contact with the $^3$He flow. The excellent sample
thermalization obtained is this way has already proven to be
essential for NMR experiments on molecular magnets performed in
the same refrigerator down to $T \simeq 15$ mK, whose success
depends crucially on the efficient cooling of the nuclear spins
\cite{morello04CM}.

The temperature inside the mixing chamber is monitored by
measuring, with a Picowatt AVS-47 bridge, the resistance of a
Speer carbon thermometer placed in the outer part of the top
Araldite pot. By adding an extra Speer thermometer (calibrated
against the previous one) in the bottom of the Kapton tail, we
have verified that the temperature is very uniform along the whole
chamber (except in the presence of sudden heat pulses): even at
the lowest $T$, the mismatch between the measured values is
typically $\lesssim 0.5$ mK. During normal operation, the
thermometer in the bottom of the tail is obviously omitted, to
avoid its contribution to the measured magnetic signal. A Leiden
Cryogenics Triple Current Source is used to apply heating currents
to a manganin wire, anti-inductively wound and glued around a
copper joint just above the $^3$He inlet in the mixing chamber. In
this way we can heat the incoming $^3$He stream and uniformly
increase the mixing chamber temperature.

For the $^3$He circulation we employ an oil-free pumping system,
consisting of a Roots booster pump (Edwards EH500) with a pumping
speed of 500 m$^3$/h, backed by two 10 m$^3$/h dry scroll pumps
(Edwards XDS10) in parallel. The main pumping line is a
$\varnothing$ 100 mm solid tube, fixed at the pump side with a
flexible rubber joint that allows for an inclination of a few
degrees, and is connected to the head of the fridge by a ``T''
piece with two extra rubber bellows, to reduce the vibrations
transmitted by the pumping system. In this configuration, the
system reaches a base temperature of 9 mK. The typical $^3$He
circulation rate at the base temperature is $\dot{n} \sim 350$
$\mu$mol/s, and the cooling power at 100 mK is $\dot{Q} \sim 150$
$\mu$W. When the Kapton tail is replaced by a flat plug, $\dot{Q}$
can be increased up to 700 $\mu$W @ 100 mK and $\dot{n} \sim 1200$
$\mu$mol/s by applying extra heat to the still; with the tail in
place it's more difficult to increase the circulation rate, and
$\dot{Q}$ @ 100 mK hardly exceeds 250 $\mu$W.

\section{Principles of SQUID magnetometery}

A SQUID (Superconducting QUantum Interference Device) is basically
an ultra-sensitive flux-to-voltage converter, that exploits the
peculiar quantum properties of closed superconducting circuits. A
detailed description of its working principle can be found in
textbooks \cite{gallopB,lounasmaaB}, or in \S 2.4.1 of Ref.
\cite{morelloT}. Here we just recall that a \textit{dc}-SQUID,
like the one used in our system, consists of a superconducting
ring interrupted by two Josephson junctions, $J_1$ and $J_2$, each
characterized by a critical current $I_c$, which represents the
maximum current that may flow through the junction without
dissipation. The \textit{dc}-SQUID is operated by biasing the
junctions $J_1$ and $J_2$ with a current $I_{\mathrm{bias}}>I_c$,
such that a voltage $V$ develops across them (Fig. 3). The
essential feature is that the voltage can be modulated by an
applied magnetic flux $\Phi$, since the critical current $I_c$
also depends on $\Phi$. One finds that the screening current $I_s$
that circulates in the SQUID ring is related to the critical
current $I_c$ of the junctions (assumed to be identical) by:
\begin{eqnarray}
I_s = \frac{I_c}{2} \left[ \sin \varphi' - \sin \left( \varphi' -
2\pi \frac{\Phi_{\mathrm{ext}} + LI_s}{\Phi_0} \right) \right],
\label{IsDC}
\end{eqnarray}
where $\Phi_0 = h/2e = 2.07 \times 10^{-15}$ Wb is the flux
quantum, $L$ is the self-inductance of the ring and $\varphi'$ is
the shift in the phase of the superconducting order parameter
which develops across each junction. $I_c$ is obtained by
maximizing $I_s$, yielding the periodic form of
$I_c(\Phi_{\mathrm{ext}})$ shown in Fig. 3. By properly choosing
the bias conditions, the \textit{dc}-SQUID operates therefore as a
flux-to-voltage converter, where the external flux can be applied
by injecting a current in the input coil. Notice that a small
fraction of a flux quantum can produce voltage changes of the
order of millivolts! In the practice a \textit{dc}-SQUID is used
in feedback mode by employing a so-called Flux-Lock Loop (FLL),
i.e. adding a feedback coil that produces a compensating flux such
that $V = const$. This increases the accuracy and the dynamic
range of the measurement, and allows to implement noise-reducing
detection schemes.

\begin{figure}[t]
\includegraphics[width=8.5cm]{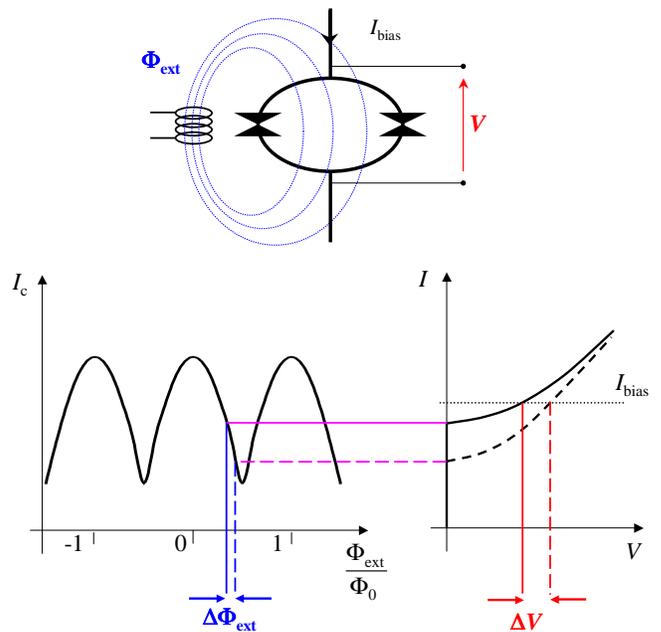}
\caption{\label{VvsPhi} Working principle of a \textit{dc}-SQUID
as flux-to-voltage converter.}
\end{figure}

To use a \textit{dc}-SQUID in an actual magnetometer, it is still
necessary (except in a rather radical design like the
``microSQUID'' \cite{wernsdorfer00JAP(SQUID)}) to produce a
current proportional to the magnetic moment of the sample to be
measured. Such a current can be injected in the input coil to
produce a flux that is coupled to the SQUID ring by the mutual
inductance $\mathcal{M}$. Typically, the input current is obtained
by constructing a closed superconducting circuit which includes
the SQUID's input coil on one side, and terminates with a pick-up
coil on the other side. Once the circuit has been cooled down
below the superconducting critical temperature, the enclosed flux
is constant. Any change in the magnetic permeability of the
circuit, for instance due to the sample, will result in a
screening current that, while keeping the total flux constant,
produces the required flux in the input coil. In our case, the
change in permeability of the pick-up circuit is obtained by
vertically moving the \emph{whole} dilution refrigerator, whose
tail, that contains the sample, is inserted in the pick-up coil.

\begin{figure}[t]
\includegraphics[width=8.5cm]{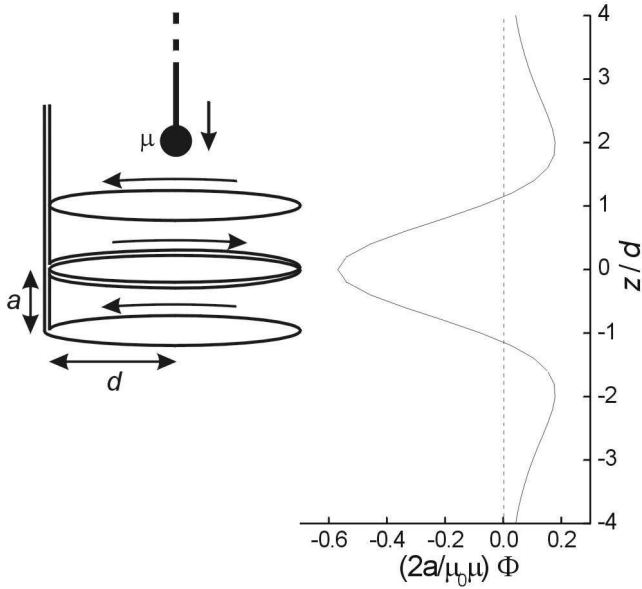}
\caption{\label{gradiometer} A second-order gradiometer in
Helmoltz geometry ($a = 2d$) and the magnetic flux induced by a
dipole moving along $z$. Notice that the side peaks in $\Phi(z)$
do not coincide with the positions of the external coils. The
factor $a$ in the $x$-axis scale is obtained by using the second
form of Eq. (\ref{fdiz}).}
\end{figure}

Because of the high sensitivity, it is essential to make sure that
no sources of flux other than the sample may couple with the
SQUID. This can be done by employing a gradiometer, which in its
simplest form consists of two coils wound in opposite direction,
so that the flux produced by any uniform magnetic field cancels
out, and only the gradient of $B$ can be detected (first-order
gradiometer). A further improvement is the second-order
gradiometer, shown in Fig. \ref{gradiometer}, which is obtained by
inserting a coil with $2N$ windings between two coils with $N$
turns each, wound opposite to the central one. This design
eliminates also the effects of linear field gradients.
Furthermore, according to the reciprocity principle
\cite{mallinson66JAP}, the flux $\Phi$ produced in a coil of
arbitrary geometry by a magnetic moment $\mu$ at position
$\vec{r}$ is related to the field $\vec{B}(\vec{r})$ produced by
the same coil in $\vec{r}$ when carrying a current $I$ such that:
\begin{eqnarray}
\vec{B}\cdot \vec{\mu} = \Phi I. \label{recip}
\end{eqnarray}
The field produced by a single-loop coil is equivalent to the
field of a magnetic dipole, whereas a first-order gradiometer is a
magnetic quadrupole, and a second-order is an octupole, so that
the fields they produce vary in space like $r^{-3}$, $r^{-4}$ and
$r^{-5}$, respectively. From Eq.(\ref{recip}) it is clear that a
second-order gradiometer is the least sensitive to magnetic fields
produced outside of the coil system.

The magnetic moment of the sample can be detected by moving it
through the pick-up coil. The enclosed flux is easily obtained
from the flux induced by a dipole with magnetic moment $\vec{\mu}
\parallel \vec{z}$ at a position $z$ along the axis in a loop of radius
$a$ placed at $z=z_0$:
\begin{eqnarray}
\Phi_{\mathrm{loop}}(z) = \frac{\mu_0}{2} f(z-z_0) \mu,\\
f(z-z_0) =  \frac{a^2}{[a^2 + (z-z_0)^2]^{3/2}} = \nonumber \\
\frac{1}{a} \left[1 +
\frac{z_0^2}{a^2}\left(\frac{z}{z_0}-1\right)^2 \right]^{-3/2}.
\label{fdiz}
\end{eqnarray}
If the upper and lower coils of the gradiometer are placed at
$z=d$ and $z=-d$, respectively, then the total picked-up flux is:
\begin{eqnarray}
\Phi_{\mathrm{pu}}(z) = N\frac{\mu_0}{2} [f(z-d) - 2f(z) + f(z+d)]
\mu.   \label{Phipu}
\end{eqnarray}
When adopting the Helmoltz geometry, $a = 2d$, the resulting
$\Phi(z)$ is shown in Fig. \ref{gradiometer}.

The picked-up flux is related to the screening current in the
pickup circuit, $I_S$, by:
\begin{eqnarray}
\Phi_{\mathrm{pu}} = (L_{\mathrm{pu}} + L_{\mathrm{leads}} +
L_{\mathrm{in}})I_S,
\end{eqnarray}
where $L_{\mathrm{pu}}$, $L_{\mathrm{leads}}$ and
$L_{\mathrm{in}}$ are the inductances of the pick-up coil, the
leads and the SQUID input coil, respectively. The flux at the
SQUID sensor $\Phi_{\mathrm{SQUID}}$ is thus given by:
\begin{eqnarray}
\Phi_{\mathrm{SQUID}} = \mathcal{M}I_s = f_{\mathrm{tr}} \Phi_{\mathrm{pu}}, \nonumber\\
f_{\mathrm{tr}} = \frac{\mathcal{M}}{L_{\mathrm{pu}} +
L_{\mathrm{leads}} + L_{\mathrm{in}}}, \label{phisquid}
\end{eqnarray}
where $f_{\mathrm{tr}}$ is the flux-transfer ratio
\cite{claassen75JAP}.

\section{Design and construction of the pick-up circuit}
\label{SQUIDconstr}

The circuitry for our SQUID magnetometer is, with a few additions,
based on the principles discussed above. The niobium
\textit{dc}-SQUID sensor is part of a Conductus LTS iMAG system,
which includes a FLL circuitry to be placed just outside the
cryostat, and is connected to the SQUID controller by a hybrid
optical-electrical cable. To couple a magnetic signal to the SQUID
we constructed a second-order gradiometer by winding a
$\varnothing$ $100$ $\mu$m NbTi wire on a brass coil-holder, to be
inserted in the bore of the Oxford Instruments 9 T NbTi
superconducting magnet. The gradiometer coils have $\varnothing$
$36$ mm ($a=18$ mm) and $d=9$ mm. The choice of the diameter is
constrained by the $\varnothing$ $34$ mm vacuum can of the
refrigerator that contains the sample. The leads of the pick-up
circuit, which are tightly twisted and shielded by a Nb capillary,
are screwed onto the input pads of the SQUID to obtain a closed
superconducting circuit. The input inductance of the SQUID sensor
is $L_{\mathrm{in}} = 600$ nH; since $L_{\mathrm{leads}} \sim 200$
nH and, given the dimensions, $L_{\mathrm{pu}} \sim
L_{\mathrm{in}} + L_{\mathrm{leads}}$ already with $N=1$, it
follows from Eq. (\ref{phisquid}) that the most convenient choice
of windings for the gradiometer is 1-2-1 (recall that
$\Phi_{\mathrm{pu}} \propto N$ but $L_{\mathrm{pu}} \propto N^2$).
The mutual inductance between SQUID ring and input coil is
$\mathcal{M} = 10$ nH, which means that $f_{\mathrm{tr}} \sim
1/200$.

The FLL included in the Conductus electronics is able to
compensate for $500$ $\Phi_0$ at the SQUID ring, which means that
a maximum flux $\Phi_{\mathrm{pu}}^{\mathrm{(max)}} =
500/f_{\mathrm{tr}} \sim 10^5$ $\Phi_0$ can be picked up by the
gradiometer without saturating the system. The maximum measurable
magnetic moment $\mu^{\mathrm{(max)}}$ is therefore [cf. Eq.
(\ref{Phipu}) and Fig. \ref{gradiometer}]:
\begin{eqnarray}
\mu^{\mathrm{(max)}} \approx \frac{2a}{0.6 \mu_0}
\Phi_{\mathrm{pu}}^{\mathrm{(max)}} \sim 10^{18}\mu_B \sim 10^{-5}
\mathrm{\ Am^2}.
\end{eqnarray}
In order to extend the dynamic range, we have built an extra flux
transformer on the pick-up circuit, which allows to introduce a
magnetic flux from the outside to compensate for the flux induced
by the sample. By using the SQUID as a null-meter, there is the
extra advantage that no current circulates in the pick-up circuit,
thus the field at the sample is precisely the field produced by
the magnet, without the extra field that would be produced by a
current in the gradiometer. The flux transformer consists of 16
turns of NbTi wire, wound on top of one loop of the pick-up wire
and shielded by a closed lead box. In the same box we tightly
glued the pick-up leads on top of a 100 $\Omega$ chip resistor,
which is used to locally heat the circuit above the
superconducting $T_c$ and eliminate the trapped flux. The flux
transformer is fed by a Keithley 220 current source, whereas the
heater is operated by one of the three current sources in the
Leiden Cryogenics Triple Current Source used for the refrigerator.
Both sources are controlled by the software described in Sect.
\ref{automation}.

Despite the efforts to isolate the system from external
vibrations, the powerful pumps for the dilution refrigerator still
provide a non-negligible mechanical noise. In particular, the
Roots pump has a vibration spectrum with a lowest peak at $\sim
25$ Hz. To prevent the possible flux changes induced by such
vibrations when coupled to a magnetic field, we added an extra
low-pass filter on the pick-up circuit\cite{meschkeT} in the form
of a 17 mm long $\varnothing$ $0.3$ mm copper wire in parallel
with the pick-up leads. At $T=4.2$ K such a wire has a resistance
$R \simeq 11$ $\mu \Omega$, which together with the input
inductance $L_{\mathrm{in}} = 600$ nH of the SQUID yields a cutoff
frequency $f_{l.p.} = (1/2\pi) (R/L_{\mathrm{in}}) \simeq 3$ Hz.
The filter is contained in a separate Pb-shielded box inserted
between the flux transformer and the SQUID. The pick-up leads and
the copper wire are contacted via Nb pads.

\begin{figure}[t]
\includegraphics[width=8.5cm]{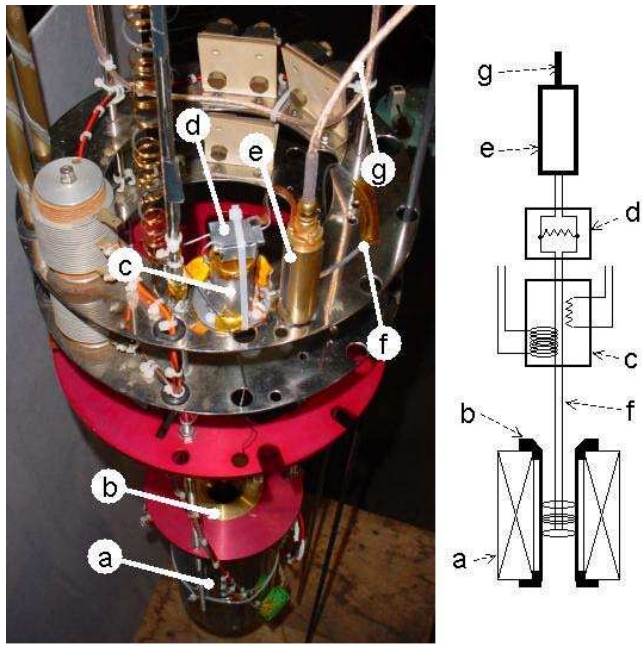}
\caption{\label{SQUIDcirc} Picture and scheme of the SQUID
circuitry. (a) Superconducting magnet. (b) Coil holder. (c) Flux
transformer and heater in lead shield. (d) 3 Hz lead shielded
low-pass filter. (e) SQUID sensor. (f) Pick-up leads with Nb
capillary shield. (g) SQUID cryocable.}
\end{figure}

A picture of the SQUID circuitry is shown in Fig. \ref{SQUIDcirc}.
The SQUID, the filter and the flux transformer are mounted on a
radiation shield of the magnet hanging. The whole system is
inserted in a 90 liter helium cryostat produced by Kadel
Engineering. The magnet hanging is stabilized by triangular
phosphor-bronze springs (not shown in Fig. \ref{SQUIDcirc}) that
press against the inner walls of the cryostat.

\section{Vertical movement} \label{verticalmov}

The movement of the sample through the gradiometer is obtained by
lifting the whole dilution refrigerator. For this purpose, the
refrigerator is fixed on a movable flange, while the top of the
cryostat is closed by a rubber bellow. On the flange we screwed
the nuts of three recirculating balls screws (SKF SN3 $20\times
5\mathrm{R}$), which allow a very smooth displacement of the nut
by turning the screw. The base of each screw is mounted on ball
bearings and is fitted with a gearwheel. The gearwheels are
connected by a toothed belt (Brecoflex 16 T5 / 1400) driven by a
three-phase AC servomotor (SEW DFY71S B TH 2.5 Nm) that can exert
a torque up to 2.5 Nm. In this way, the rotation of the servomotor
is converted into the vertical movement of the dilution
refrigerator insert. The movement is so smooth that the consequent
vibrations are hardly perceptible and do not exceed the vibrations
due to the $^3$He pumping system.

The motor is driven by a servo-regulator (SEW MDS60A0015-503-4-00)
that can be controlled by a computer. For safety reasons an
electromagnetic brake is fitted, that blocks the motor in case of
power failure. In addition, a set of switches is mounted along one
of the pillars that support the screws: when the flange reaches
the highest or the lowest allowed position, the switches force the
motor to brake, independently of the software instructions. The
maximum allowed vertical displacement is 10 cm.

\begin{figure}[t]
\includegraphics[width=8.5cm]{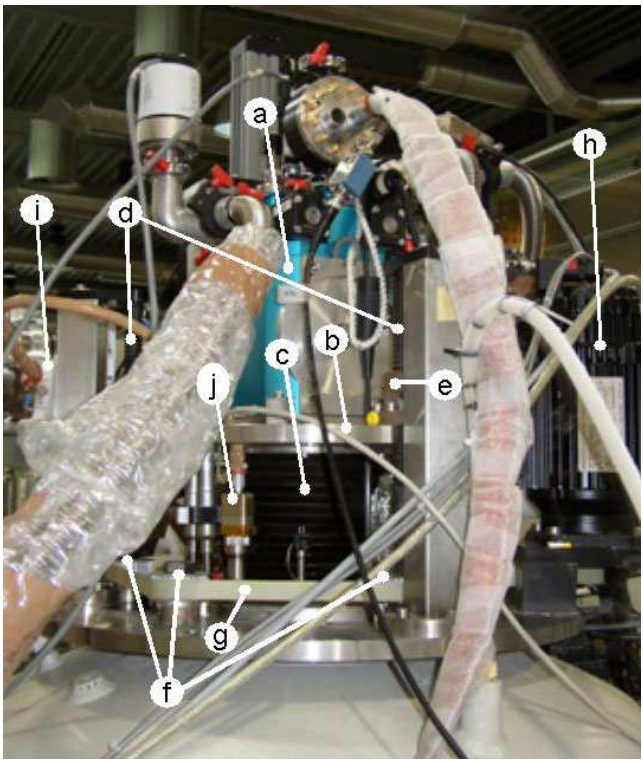}
\caption{\label{setuptop} Picture of the top of the cryostat, with
the elements for the vertical movement. (a) Head of the dilution
refrigerator. (b) Movable flange. (c) Rubber bellow. (d) Screws.
(e) Nut with recirculating balls. (f) Gearwheels. (g) Belt. (h)
Servomotor. (i) SQUID FLL electronics. (j) Feedthrough to the
SQUID cryocable.}
\end{figure}

A picture of the top of the cryostat with the vertical movement
elements is shown in Fig. \ref{setuptop}.

\section{Grounding and shielding}

The shielding of the SQUID sensor and electronics is of course a
crucial issue for the successful design of a SQUID-based
magnetometer. As mentioned in \S\ref{SQUIDconstr}, the
low-temperature parts are shielded by superconducting Pb boxes or
Nb capillaries. As for the electronics outside the cryostat, the
Conductus iMAG system provides hybrid optical-electrical cables
for the communication between the SQUID controller and the FLL
electronics on top of the cryostat, but in the cable that connects
the FLL to the SQUID sensor, the ground is used as return line for
the signals. This means that \emph{any ground loop} involving the
SQUID electronics will \emph{completely destroy} the functionality
of the system. Obviously, all the outer metallic parts in the
system (e.g. the case of the SQUID sensor, the vacuum feedthrough
for the cryocable, etc.) are connected to the same electrical
ground, including the GPIB communication terminals.

The best way to avoid ground loops in the SQUID system would be to
ground the SQUID sensor and its cable at the cryostat (and take
care that it remains a very clean ground), and connect the
controller to an isolation transformer \cite{vleemingT}. This
method is not practicable when communication with the computer via
the GPIB bus is needed, and the same computer is connected to
another instrument that requires a common ground with the cryostat
(this is the case for the Picowatt AVS-47 resistance bridge). The
only choice is therefore to ground the controller at its power
cord and \emph{float the whole SQUID circuitry}, all the way to
the SQUID sensor inside the cryostat and the shields of the
pick-up circuit (which must be connected to the SQUID ground).
This is already a rather cumbersome operation, but it's not yet
sufficient. We found out that, in this configuration, the FLL
electronics and the room-temperature cables around it are not
enough shielded from the electromagnetic interference generated by
the motor during the vertical movement of the dilution
refrigerator (we obviously took care that the motor does not touch
the cryostat ground). This sort of interference does not
annihilate the functionality of the SQUID like a ground loop would
do, nor does it simply induce an increase of the instrumental
noise: the effect of bad shielding is that the SQUID system
behaves as if it were connected to a non-superconducting,
inductive circuit. This fault is hard to discover, since the
inductive behavior of the pick-up system is one of the most
expectable failures, which can be due to any weakening of the
superconductivity in the circuit, for instance because of a bad
contact on the SQUID input pads. The full functionality of the
magnetometer was reached by enclosing the FLL and the
room-temperature SQUID cables into an extra copper shield,
grounded at the cryostat but separated from the SQUID ground.

\section{Automation} \label{automation}
\hyphenation{MultiVU}

\begin{figure}[t]
\includegraphics[width=8.5cm]{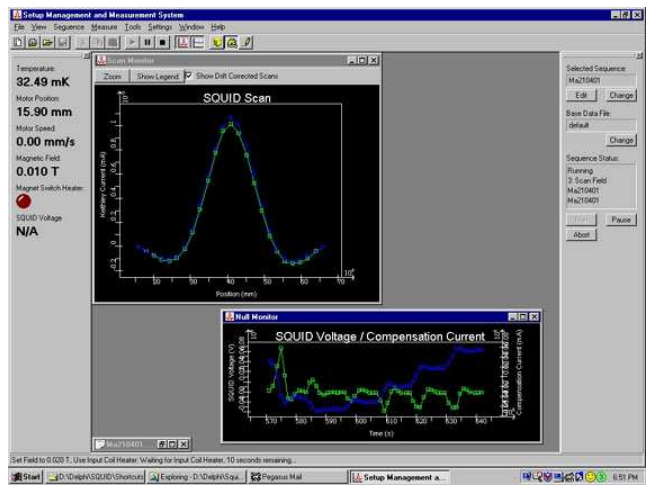}
\caption{A screenshot of the application's user interface.
Information about temperature, vertical position, applied magnetic
field, SQUID voltage and status of the program is constantly
provided in the workspace. Additional windows are open, in this
case, to monitor the result of the measurement and the nulling of
the SQUID voltage.} \label{FigUI}
\end{figure}

The automation of the measurement system is achieved by connecting
to a PC the iMAG SQUID controller, the Keithley 220 current
course, the LC Triple Current Source, the Picowatt AVS 47
resistance bridge, and the SEW servo-regulator. The PC runs a
home-written software, which provides an application with a single
but flexible Graphical User Interface (GUI), containing all the
required functionality: setup management, monitoring, measurement,
automated sequence execution and data analysis. The application is
written in Borland Delphi because of its power, ease of use, good
documentation and rapid GUI development possibilities. The Windows
user interface is primarily inspired upon the Quantum Design's
MultiVU application, included with their commercial Magnetic
Property Measurement System (MPMS). This has the advantage of
offering an interface that looks familiar to many users. The GUI
consists of a Multiple Document Interface (MDI) with among other
things a menu, a toolbar, a status bar, and removable panels. The
MDI workspace is where the actual work is done in several
subwindows, as shown in Fig. \ref{FigUI}.

A set of tools are available for e.g. positioning the sample,
setting a magnetic field, monitoring the temperature and
performing ``zero field oscillations'', i.e. oscillations of the
applied field which are used to reduce the trapped flux in the
magnet coil.

In order to measure the magnetic moment of a sample, the response
of the SQUID is measured while vertically moving the sample (i.e.
the whole dilution refrigerator insert) along the gradiometer, as
illustrated in Fig. \ref{gradiometer}. The sample can be
positioned in two ways, using the so-called \textit{continuous} or
\textit{hold} mode. In the continuous mode, the sample is
positioned at the start of the trajectory and then set into a
continuous motion. During the movement, readings are made at fixed
time intervals. When the sample reaches the end of the trajectory,
the movement is stopped. In the hold mode, the sample is moved for
a short interval of time after which the motor is stopped and the
SQUID reading is done, repeating this sequence until the final
position is reached.

A reading can be done in \textit{direct} or \textit{null} mode. In
the direct mode, one reads directly the SQUID voltage, which is
proportional to the screening current that circulates in the
pickup system. In null mode, the compensation current in the flux
transformer is continuously being regulated by the software, using
a proportional feedback controller which runs in the background,
until the SQUID voltage is zero. A reading in null mode consists
therefore of the current necessary to null the SQUID voltage. At
the start of each measurement (``scan''), the sample is first
accurately positioned at the starting point and then, if used, the
nulling process is started. The measurement will then start and
continue until the end point is reached. Typically two scans are
done, an upward and a downward scan; the results of the two scans
are corrected for the drift of the SQUID voltage, averaged and
fitted to obtain the dipole moment.

\begin{figure}[t]
\includegraphics[width=8.5cm]{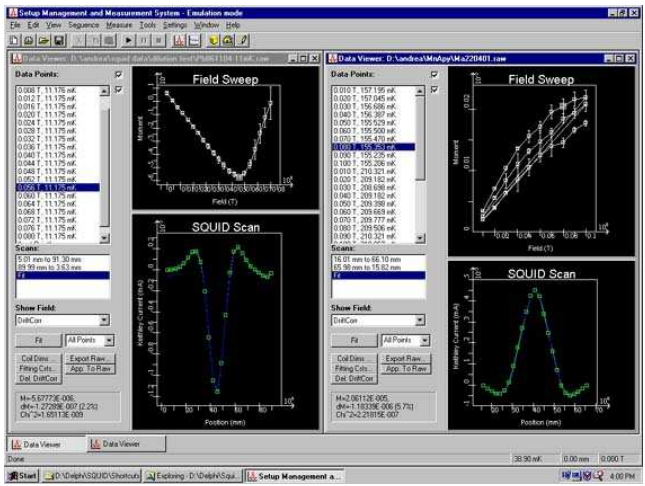}
\caption{A screenshot of the data viewer. On the left-hand side we
analyzed the field dependence of the magnetization of a Pb sample
between 0 and 80 mT at $T=11$ mK. On the right-hand side, a
sequence of magnetization curves of MnApy (see text) between 0 and
100 mT at four different temperatures.} \label{FigDataViewer}
\end{figure}

Measurements can be managed, viewed and fitted using the built-in
data viewer: an example is shown in Fig. \ref{FigDataViewer}. The
data viewer consists of a set of controls and two graphs, one
containing the fitted dipole moment of multiple scans and the
other containing the single fitted scans.

In the end, the fitted dipole moment obtained from a set of scans
which have been interpolated, averaged and fitted, represents one
data point at a certain temperature and applied magnetic field. In
general, we are interested in measuring the sample magnetization
at several values of temperature and/or field. To do this
automatically, an elementary scripting language has been
implemented very similar to the way done in the MPMS MultiVU
application. There are commands for, among other things, changing
the sample position, setting a temperature, setting a field,
heating and retuning the SQUID sensor, performing zero field
oscillations and running a subsequence. All these commands can be
nested in for-loop structures for measuring as a function of
temperature or field. The sequence files are edited using a
built-in sequence editor which supports the most common editing
functions such as cut, copy, paste and undo. When a sequence is
run, it can be paused, resumed, aborted and the execution status
can be monitored.

As for the implementation, the application consists of a main core
structure and modular components called ``applets''. The core
structure implements routines for the starting and stopping of the
program, user interface, device initialization and communication,
and the necessary menu items, tool buttons and shortcuts for
starting the various components of the application. The applets
are started from the main program and do all the work involving
device communication, e.g. provide a tool for magnet management,
monitor the status of the setup, null the SQUID using the FLL, do
a set of scans, or execute a sequence. The applets which
communicate with devices run in separate parallel threads to
ensure communication without sacrificing performance of the main
program. There are, for example, typically four threads running
during a null-mode measurement:

\begin{itemize}\itemsep-0.1cm
\item The main GUI thread

\item A thread which monitors and displays the mixing chamber temperature, vertical sample
position, magnet status and SQUID voltage. This thread
communicates with devices not used by other applets.

\item The measurement thread, which communicates with the servo regulator and the SQUID (in direct
mode)

\item The nuller thread which communicates with the SQUID and Keithley 220 current source

\end{itemize}

When an applet obtains information from a device, e.g. servomotor
position, magnetic field or SQUID voltage, this information is
shared with the other applets so that all applets stay informed
about the global status of the system.

The applets communicate with the devices through a central
abstract layer which encapsulates the device specific hardware
communication and protocols. This central layer also makes it
possible to start and stop the devices while the application is
running. All the device's functionality can additionally be
simulated, which was very useful during the implementation phase:
most of the application was written without using the experimental
setup.

\section{Performance}

\begin{figure}[t]
\includegraphics[width=8.5cm]{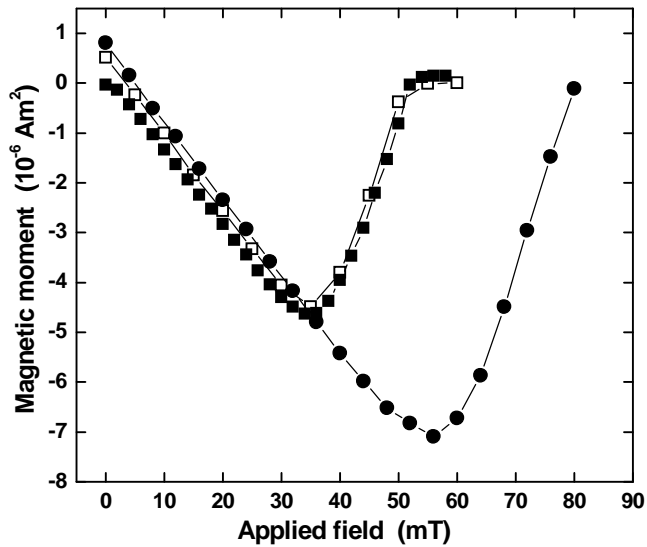}
\caption{The magnetic moment of a small Pb grain, as measured in
our setup (full symbols) and in a Quantum Design MPMS-XL
magnetometer (open symbols). Squares: $T=4.5$ K, circles: $T=11$
mK.} \label{lead}
\end{figure}

In order to check the performance of our instrument, we first
measured the magnetization of a small lead grain in a Quantum
Design MPMS-XL magnetomter, then we mounted the sample, with the
same sample holder, inside the mixing chamber of the dilution
refrigerator. The results are shown in Fig. \ref{lead}: at $T=4.5$
K, the data in our setup and in the MPMS-XL can be accurately
superimposed, provided the correct calibration factor is chosen.
By cooling down to $T=11$ mK, we still find the same slope of
$M(H)$ and an expectedly higher critical field.

As an example of \textit{dc}-susceptibility experiment, we show in
Fig. 10 a dataset obtained by measuring 1.9 mg of the paramagnetic
salt manganese antipyrine
Mn(C$_{11}$H$_{12}$ON$_2$)$_6$(ClO$_4$)$_2$ (MnApy), in a constant
applied field $\mu_0 H_{\mathrm{a}} = 2$ mT while slowly cooling
down the refrigerator. The paramagnetic centers in MnApy are the
Mn$^{2+}$ ions, with spin $S = 5/2$ and gyromagnetic factor $g =
2.00$. In this sort of measurements we typically use the
``continuous + direct'' mode, which allows to perform the scans
rather quickly ($\sim 1$ min) as compared to the timescale of the
temperature changes. The volume susceptibility $\chi = M/H = \mu
\rho / (mH)$ ($m$ is the mass of the sample and $\rho = 1.361
\times 10^3$ kg/m$^3$ is the density) can be obtained from the
measured magnetic moment $\mu$ and compared with the expected
behavior of a Curie paramagnet $\chi = C/T$, where $C = \mu_0
\mathcal{N} g^2 \mu_B^2 S(S+1) / (3 k_B)$ is the Curie constant
and $\mathcal{N} = 5.927 \times 10^{26}$ m$^{-3}$ is the number of
spins per unit volume. In view of the high value of $\chi$
obtained at very low temperature, the comparison is meaningful
only after having corrected the raw data for demagnetizing
effects:
\begin{eqnarray}
\chi_{\mathrm{corr}} = \frac{\chi_{\mathrm{raw}}} {1 - D
\chi_{\mathrm{raw}}},
\end{eqnarray}
where $D$ is the demagnetizing factor.

Since the sample is in powder form, $D$ will be an effective
demagnetizing factor appropriate for the container (in this case a
cylindrical capsule) filled with the grains. This is a notoriously
complex problem that can only be solved in an approximate way. In
a first step the relation should be found between the applied
field $H_{\mathrm{a}}$ and the local field $H_{\mathrm{cont}}$
acting on a grain in the container. The dipolar contributions from
the other grains can be divided in a shape-dependent part given by
the demagnetizing factor $D_{\mathrm{cont}}$ of the container, and
a ``structural'' part given by the packing of the grains. In a
second step the relation between $H_{\mathrm{cont}}$ and the local
field $H_{\mathrm{loc}}$ inside a grain should be found. The
difference between $H_{\mathrm{cont}}$ and $H_{\mathrm{loc}}$ will
now be due to the shape-dependent contribution related to the
demagnetizing factor $D_{\mathrm{grain}}$ of the grain and the
structural part arising from the crystal structure of the
material.

In view of all the uncertainties involved, the simplest solution
for the present purpose appears to assume all grains to have
identical shape and to adopt, for both steps, the Lorentz
approximation to estimate the contributions depending on the
structural aspects. Thereby we shall also neglect the
contributions from the dipoles inside the spheres. One then
obtains:

\begin{eqnarray}
D = D_{\mathrm{grain}} - \frac{1}{3} + f\left(D_{\mathrm{cont}} -
\frac{1}{3}\right) = f\left(D_{\mathrm{cont}} -
\frac{1}{3}\right), \label{demagfact}
\end{eqnarray}
where $f$ is the filling fraction (volume) of the container and
where the last equality follows from the assumption
$D_{\mathrm{grain}}=1/3$, i.e. we assume all grains to have a
spherical shape. Given the shape of the sample holder, we estimate
the demagnetizing factor of the container $D_{\mathrm{cont}}
\simeq 0.7$, which yields $D \simeq 0.37 f$. Taking for the
filling factor $f \simeq 0.5$ (ideally packed spheres would yield
$f \simeq 0.64$), we obtain the corrected volume susceptibility
shown in Fig. 10. $\chi_{\mathrm{corr}}$ closely follows the
calculated ideal Curie law, except for some small deviations below
$\sim 0.2$ K. Such deviations are better appreciated when plotting
$1/\chi_{\mathrm{corr}}$ \textit{vs} $T$, as shown in the inset of
Fig. 10. As it appears, the susceptibility remains always slightly
lower than the value for an ideal Curie paramagnet. This can be
attributed to a small crystal field anisotropy of the Mn$^{2+}$
electron spins, which causes a reduction of the effective Curie
constant at very low temperatures due to the depopulation of the
excited energy levels in the zero-field split manifold
\cite{algra77PhyB}. Another reason for the deviation may be an
(antiferromagnetic) ordering of the moments, e.g. due to the
dipolar interactions between them. As an estimate of the
interaction energy, $U$, we may consider the dipolar coupling
between two Mn$^{2+}$ spins $5/2$ at a distance $r \simeq 1$ nm in
the MnApy compound. One then obtains $U = (\mu_0/4\pi)(g \mu_B
S)^2/(k_B r^3) \simeq 15$ mK. Although this estimate is very
rough, it is an indication that such an ordering phenomenon may
not be excluded.

The uncertainty in the choice of $D$ makes it rather delicate to
analyze the data in more detail, but for the purpose of the
present discussion we may state that the thermalization of the
sample does not pose any special problem down to our lowest
achievable temperature.

\begin{figure}[t]
\includegraphics[width=8.5cm]{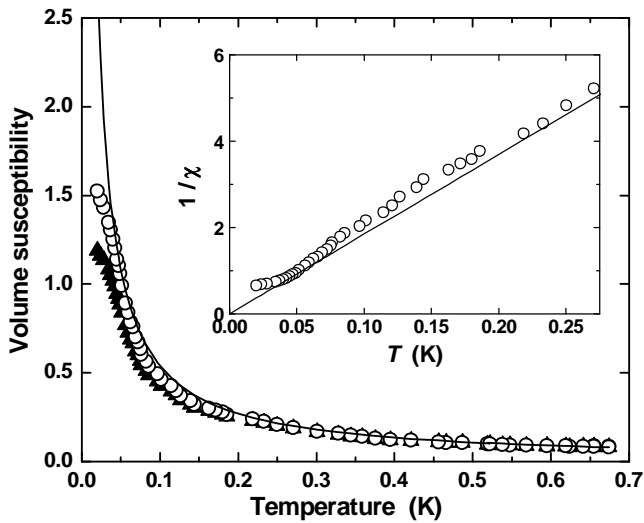}
\caption{Full triangles: measured \textit{dc}-susceptibility of
MnApy in an applied field $\mu_0 H_{\mathrm{a}} = 2$ mT. Open
circles: data corrected for the demagnetizing factor (see text).
Solid line: calculated susceptibility according to the Curie's
law. Inset: inverse of the corrected susceptibility (open
circles), fitted by the Curie-Weiss law (solid line), yielding
$\Theta \simeq 14$ mK.} \label{mnapy}
\end{figure}

The typical noise at the SQUID in normal operation, i.e. while all
the pumps are running and the dilution refrigerator insert is
being moved, is $\sim 1$ mV, which translates into an equivalent
magnetic noise $\sim 10^{-9}$ Am$^2$ ($10^{-6}$ emu). Measuring a
magnetic moment $\sim 10^{-4}$ Am$^2$ (0.1 emu) in ``null mode''
requires a compensation current $\sim 30$ mA, which poses no
problems to the compensation circuitry; this means that our system
can easily cover a dynamic range of five orders of magnitude in
magnetic moment.

As for the maximum applicable magnetic field, although the magnet
itself can produce 9 T, the SQUID circuitry tends to become
unstable above $\sim 0.2$ T. This a rather typical value when no
special cautions are taken to stabilize the field and screen the
SQUID circuitry from the stray fields: in specialized
systems\cite{genna91RSI,paulsenB}, operation up to 8 T has been
achieved by designing a dedicated magnet with large bore and by
adding a NbTi superconducting shield around the sample- and pickup
coils- space. We recall that our system is not dedicated uniquely
to SQUID magnetometery and can accommodate, for instance, NMR
experiments where the possibility of producing continuous field
sweeps is essential.

\begin{acknowledgements}
We thank J. Sese for the precious advice about SQUID sensors, K.
Siemensmeyer for the suggestions on the design of the pickup
system, T. G. Sorop for the high-$T$ measurements and the fruitful
discussions, and F. Luis for his contributions in the early stage
of the project. The technical support of A. Kuijt, E. de Kuyper,
R. Hulstman and M. Pohlkamp has been invaluable and is gratefully
acknowledged. This work is part of the research program of the
``Stichting voor Fundamenteel Onderzoek der Materie'' (FOM).
\end{acknowledgements}

\noindent $^*$ Corresponding author:

L.J. de Jongh, e-mail: dejongh@phys.leidenuniv.nl

\end{document}